\documentclass[journal]{IEEEtran}

\usepackage{color, colortbl}
\definecolor{Gray}{gray}{0.9}
\definecolor{LightCyan}{rgb}{0.88,1,1}
\usepackage[first=0,last=9]{lcg}

\usepackage[table,xcdraw]{xcolor} 

\usepackage{amssymb}
\usepackage{latexsym}
\usepackage{amsfonts}
\usepackage{amsmath,stackengine}

\usepackage{cite}
\DeclareFontFamily{OT1}{pzc}{}
\DeclareFontShape{OT1}{pzc}{m}{it}{<-> s * [1.10] pzcmi7t}{}
\DeclareMathAlphabet{\mathpzc}{OT1}{pzc}{m}{it}

\makeatletter

\newcommand*{\underarrow}{\def\@underarrow{\relax}\@ifstar{\@@underarrow}{\def\@underarrow{\hidewidth}\@@underarrow}}
\newcommand*{\@@underarrow}[2][]{\underset{\@underarrow\substack{\uparrow\if\relax\detokenize{#1}\relax\else\\#1\fi}\@underarrow}{#2}}

\newcommand*{\overarrow}{\def\@overarrow{\relax}\@ifstar{\@@overarrow}{\def\@overarrow{\hidewidth}\@@overarrow}}
\newcommand*{\@@overarrow}[2][]{\overset{\@overarrow\substack{\if\relax\detokenize{#1}\relax\else#1\\\fi\downarrow}\@overarrow}{#2}}
\makeatother

\usepackage{flushend}

\usepackage{graphicx}        
\usepackage{amsmath}
\usepackage{lipsum}
\usepackage{mathtools}
\usepackage{cuted}

\definecolor{amber}{rgb}{1.0, 0.75, 0.0}
\definecolor{almond}{rgb}{0.94, 0.87, 0.8}
\definecolor{blond}{rgb}{0.98, 0.94, 0.75}
\definecolor{cornflowerblue}{rgb}{0.39, 0.58, 0.93}
\definecolor{lavenderblue}{rgb}{0.8, 0.8, 1.0}
\definecolor{lightskyblue}{rgb}{0.53, 0.81, 0.98}
\definecolor{lime(web)(x11green)}{rgb}{0.0, 1.0, 0.0}
\definecolor{lime(colorwheel)}{rgb}{0.75, 1.0, 0.0}
\definecolor{persianpink}{rgb}{0.97, 0.5, 0.75}
\definecolor{mistyrose}{rgb}{1.0, 0.89, 0.88}
\definecolor{ticklemepink}{rgb}{0.99, 0.54, 0.67}
\definecolor{salmonpink}{rgb}{1.0, 0.57, 0.64}
\definecolor{richbrilliantlavender}{rgb}{0.95, 0.65, 1.0}
\definecolor{pink}{rgb}{1.0, 0.75, 0.8}

\ifCLASSINFOpdf
\else
\fi
 
\usepackage{xcolor}
\usepackage{multirow}
\usepackage{amsbsy,amsfonts,amsfonts,amssymb,amsmath,epsfig}
\usepackage{graphicx}
\usepackage{graphics}
\usepackage[cmintegrals]{newtxmath}

\makeatletter
\def\vhrulefill#1{\leavevmode\leaders\hrule\@height#1\hfill \kern\z@}
\makeatother

\begin{document}
%
\title{Reconfigurable Intelligent Surface Aided Spatial Media-Based Modulation}
%
%

\author{Burak Ahmet Ozden, Erdogan~Aydin, Fatih Cogen
	\thanks{B. A. Ozden is with Y{\i}ld{\i}z Technical University, Department of Computer Engineering, 34220, Davutpasa, Istanbul, Turkey 
 (e-mail: bozden@yildiz.edu.tr).}

	\thanks{E. Aydin is with Istanbul Medeniye University, Department of Electrical and Electronics Engineering, 34857, Uskudar, Istanbul, Turkey (e-mail: erdogan.aydin@medeniyet.edu.tr) (Corresponding author: Erdogan Aydin.)}

 \thanks{F. Cogen is with Turkish-German University, Department of Mechatronics Engineering, Istanbul, Turkey 
 (e-mail: cogen@tau.edu.tr).}
}

\maketitle

\begin{abstract}
The demands for high data rate, reliability, high energy efficiency, high spectral efficiency, and low latency communication have been increasing rapidly.  For this reason, communication models that use limited resources in the best way, allow fast data transmission, and increase performance has become very important.  In this work, a novel high energy  and spectral efficient reconfigurable intelligent surface aided spatial media-based modulation system, called RIS-SMBM, is proposed for Rayleigh fading channels. In addition to the bits carried in the $M$-QAM symbol, while media-based modulation (MBM) provides data bits to be carried in the indices of different channels according to the radio frequency (RF) mirrors are on or off, spatial modulation (SM) provides data bits to be carried in the indices of the transmit antennas. By combining these two modulation schemes, the spectral efficiency increases considerably since the amount of information transmitted in the same time interval is substantially increased. The optimal maximum-likelihood (ML) detector and the enhanced low-complexity (ELC) detector for the RIS-SMBM system are proposed. The ELC detector achieves near ML performance while reducing the complexity of the optimal ML detector for the proposed RIS-SMBM system. We analyze the average bit error rate (ABER), throughput, complexity, and energy efficiency for the RIS-SMBM scheme and verify the analytical results with Monte Carlo simulations. It has been observed that the proposed system provides better error performance as well as providing  higher spectral and energy efficiency than benchmark systems.

\end{abstract}

\begin{IEEEkeywords}
Media-based modulation, index modulation,  spatial modulation,  spatial media-based modulation, reconfigurable intelligent surface, and MIMO systems.
\end{IEEEkeywords}

%
\IEEEpeerreviewmaketitle

\vspace{-1em}
\section{Introduction}
%
%
%
%
\IEEEPARstart{S}{ince} advanced communication technologies such as 5G and 6G have been mentioned in the communication literature, and systems that can be used in integration with these technologies have become much more essential. These systems focus on features such as low energy consumption, high bandwidth, high data rate, and low latency. 
In wireless communication, obstacles in the communication path between the receiver and the transmitter and random changes in the communication channel is a severe problem that reduces the quality of the signal. Thus, many communication architectures and models have been presented to minimize the disruptive effects of the wireless channel on the signal \cite{6G_JOURNAL, Satoshi6G}.

In the wireless channel between the transmitter and the receiver, the received signal quality is considerably reduced as a result of the signal being exposed to interference effects such as reflection, scattering, and diffraction \cite{mahender2018analysis}. Multiple input multiple output (MIMO) systems, based on the technique of using multiple antennas at the transmitter and receiver, have emerged as an effective solution to combat the channel's disruptive effects \cite{AydAydlawton2004mimo,AydinSNR}. MIMO systems are used to cope with the fading effect, to achieve high spectral efficiency, and to provide better performance. Through MIMO systems, the same signal in transmitting is transmitted from more than one transmit antenna. In this way, diversity gain is obtained since the received signal is exposed to different fading effects. In another scenario, multiplexing is done by transmitting different signals from different antennas, increasing the amount of information transmitted per unit time \cite{molisch2004mimo}. Due to such important advantages, many communication architectures have been developed integrating with MIMO systems. However, the use of many antennas in MIMO systems has brought some problems. The high complexity of receiver structures, transmission delay, high power consumption, inter-channel interference (ICI), and high cost can be given as examples \cite{di2011spatial, di2013spatial}. Integrating spatial modulation (SM) and media-based modulation (MBM) schemes into MIMO systems makes a very important contribution to reducing the disadvantages of the MIMO schemes.

Index modulation (IM) schemes have a very important place in wireless communication due to the advantages they offer and their common use. The IM technique is a technique that makes it possible to carry information in the indices of the basic elements used in wireless communication systems \cite{Basarindex2016, Cogen2021, Aydn2019}. IM-based systems provide high spectral efficiency, low complexity, and high energy efficiency \cite{Mao2019, basar2017index, Aydin2019}. IM schemes have become very interesting and popular with their advantages such as high performance, high data rate and hardware simplicity \cite{Shamasundar2018_1}. One of the leading members of the IM schemes, the SM system stands out with its advantages and widespread use. The SM technique is a modulation technique based on the principle of transmitting some of the data bits through the index of the active transmit antenna of the MIMO scheme, in addition to carrying information in symbols \cite{mesleh2008spatial}. According to the symbol selected in the SM, the antenna that will be active in the transmission is determined. In this way, it is possible to carry extra information in transmit antenna indices, and extra information can be transmitted in the same time interval. The SM technique provides significant benefits such as increasing spectral and energy efficiency, reducing receiver complexity, eliminating ICI, and increasing error performance \cite{fu2014performance,mesleh2014quadrature, AydinCMRC}.

The MBM technique, which is the new and outstanding scheme of the IM family, is a digital communication method that draws attention to the benefits it provides to wireless communication schemes. In the MBM, information is carried in the indices of different channels, which are formed due to the change of the radiation pattern according to the on-off state of the radio frequency (RF) mirrors \cite{basar2019media, Aydin2021}. The spectral efficiency in the MBM directly depends on the number of RF mirrors. This means that it is possible to significantly increase energy efficiency by keeping the number of transmit antennas constant and only increasing the number of RF mirrors \cite{naresh2016media}. For this reason, MBM is seen to be much better in terms of performance when compared to other IM techniques. The RF mirrors used in the MBM system do not need hardware structures such as mixers and filters.  \cite{Shamasundar2018_2}. The fact that there is no need for complex hardware structures will bring advantages such as hardware simplicity and low cost. 

\begin{figure*}[t]
	\centering{\includegraphics[width=\textwidth]{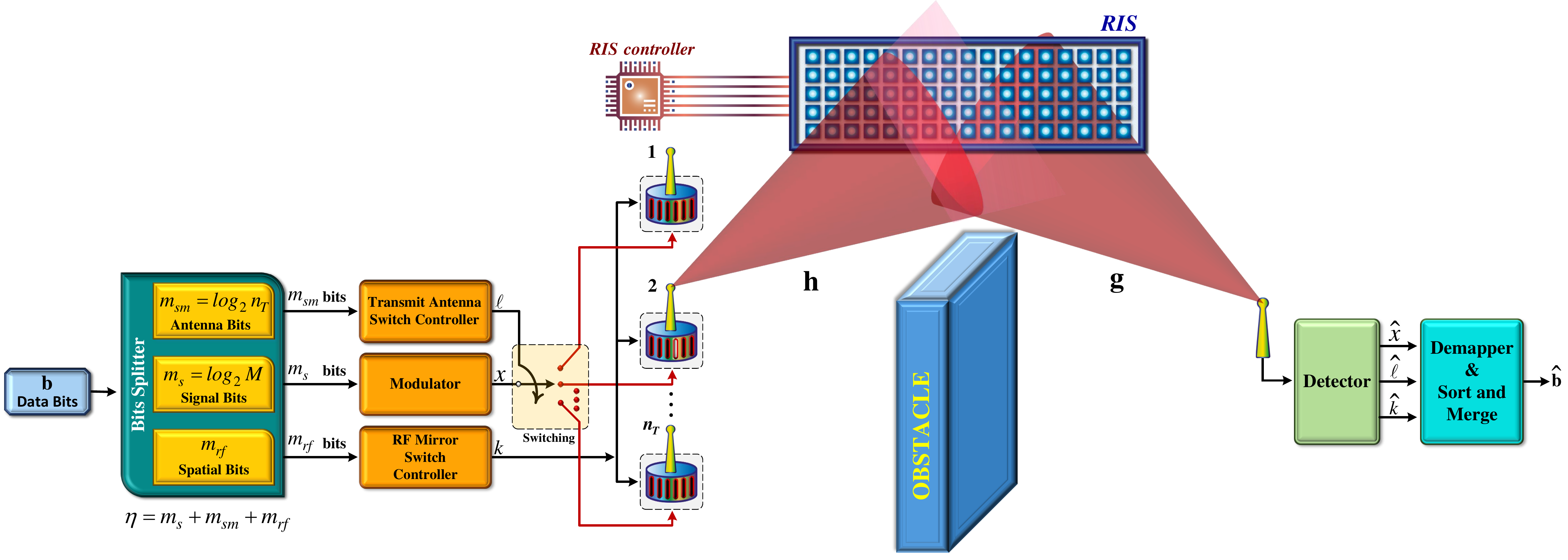}}
	\caption{The proposed RIS-SMBM scheme.}
	\label{system-model} 
	\end{figure*}

Recently, a new technology with si
gnificant communication
 capabilities has emerged called reconfigurable intelligent surface (RIS). RISs provide significant improvements in signal quality by directing the signals to hit their reflecting surfaces at certain rates. RISs are tunable smart surfaces derived from a large number of low-cost and passive elements with the ability to modify transmitted waves, which materials do not inherently have \cite{Basar2019}. Thanks to RISs, the negative effects of the channel on the signal can be decreased by controlling the scattering and reflection properties of the signals that are exposed to disruptive effects in the wireless transmission channel. The use of varactor diodes in the structure of RISs has created numerous advantages. A varactor diode is a type of diode that can take a capacitive value by functioning as a capacitor in reverse bias. Varactor diode has many advantages compared to other diodes. Compared to other diodes, it has many advantages such as producing less noise, being less costly, and is produced in very small sizes and weights. RIS is essentially a metasurface. Metasurfaces are thin, two-dimensional layers of metamaterials that can direct electromagnetic waves. The main advantages of metamaterials are their low cost, low absorption and easy integration \cite{Quevedo-Teruel2019, hsiao2017fundamentals}. RISs are distinguished from others by the many advantages they offer when compared to related technologies. RISs stand out with their serious advantages such as not needing any signal processing, power amplification, and filtering processes. Also, they are very low cost, support full-duplex and full-band transmission, and have flexible usage areas \cite{Basar2021, liu2021reconfigurable}. It is predicted that RIS technology will be widely used in advanced new generation communication networks, as it has the potential to meet the requirements brought by technological developments. Performance analysis of the SM scheme is investigated for the rapidly time-varying Rayleigh fading channels in \cite{Khattabi2019}. In \cite{Basar2020}, RIS-assisted IM schemes including RIS-aided SM (RIS-SM) and RIS-aided space shift keying (RIS-SSK) techniques are introduced. RIS-based MBM scheme is presented and RIS has been designed, aiming to demonstrate the diversity of radiation patterns in the 5G communication system in \cite{Hodge2020}. In \cite{Huang2019}, a deep learning technique is presented for the efficient wireless configuration of RISs in indoor environments. In \cite{salann2021}, over Weibull fading channels RIS-SSK/SM systems are proposed for the perfect and imperfect channel phase knowledge states at the receiver. 

\subsection{Contributions}

In this article, a novel RIS-SMBM system is considered. In the proposed technique, the data bits are mapped to antenna indices, and the indices of different channels are formed according to the on-off state of the RF mirrors in addition to conventional constellation symbols. Based on our analytical derivations, analyses, and simulation results, the contributions of our article are given item by item as follows:
\begin{enumerate}
	\item We proposed a combined system consisting of the SM and MBM schemes based RIS system, called RIS-SMBM, to improve the data rate, spectral efficiency, error performance, and energy efficiency by exploiting the multiple RF chains in the SMBM system. 
	\item Analytical expressions for the average bit error rate (ABER), complexity, energy efficiency, throughput, and data rate analysis of the RIS-SMBM scheme are derived.
	\item We propose an optimal maximum-likelihood detector (ML) and an enhanced low-complexity (ELC) detector for the RIS-SMBM system. The ELC detector achieves near ML performance while reducing the complexity of the optimal ML detector for the proposed RIS-SMBM system.

    \item The ABER performance results of the RIS-SMBM system are compared with the RIS-SM, RIS-aided quadrature SM (RIS-QSM), and RIS-empowered MBM (RIS-MBM) techniques. Simulation results prove that the RIS-SMBM system provides better error performance and high spectral efficiency than these benchmark systems while consuming less transmission energy.

    \item The analytical and simulation results of the RIS-SMBM system are obtained and it is observed that the results are very close to each other.
    
\end{enumerate}

\subsection{Organization and Notation}

The rest of this article is organized as follows. In Section II, the proposed RIS-SMBM system model is introduced. The analytical analysis of the proposed system model  is given in Section III. In Section IV, energy efficiency, throughput, data rate, and complexity analysis are presented. In Section V, simulation results, performance comparisons, and discussions are proposed. Finally, Section VI concludes the article.
	
	\textit{Notation:} The following notation is considered for this work. \textit{i})  $(\cdotp)^T$, $(\cdotp)^H$, $\big\|\cdotp \big\|_F $, $E\big[\cdotp\big]$, and $\big| \cdotp \big|$ express transpose, Hermitian, Frobenius norm, expectation, and Euclidean norm  operators, respectively. \textit{i}) Bold lower/upper case symbols represent vectors/matrices;  $\Re(.)$
	and $\Im(.)$ are the real and imaginary elements of a complex number.


\section{System Model}

The signal model, channel model, ML detector, and ELC detector for the RIS-SMBM scheme are presented in this section.

\begin{table}[t]
\centering
\caption{Mapping procedure for the proposed RIS-SMBM system}
\label{T1}
\begin{tabular}{|c||cccc|}
\hline
\rowcolor[HTML]{FFCE93} 
\hline
\hline
\textbf{}                                  & \textbf{\begin{tabular}[c]{@{}c@{}}Input bits\\ ($\eta$)\end{tabular}} & \textbf{\begin{tabular}[c]{@{}c@{}}Symbol bits\\ ($m_S$)\end{tabular}} & \textbf{\begin{tabular}[c]{@{}c@{}}rf bits\\ ($m_{rf}$)\end{tabular}} & \textbf{\begin{tabular}[c]{@{}c@{}}Antenna bits\\ ($m_{SM}$)\end{tabular}} \\ \hline\hline
\rowcolor[HTML]{FFFC9E} 
\begin{tabular}[c]{@{}c@{}}$M=4$\\ $m_{rf}=2$\\ $n_T=2$\end{tabular} & 

\begin{tabular}[c]{@{}c@{}} {[}111001{]} \\ $\eta=6$ \end{tabular}                                                             & \begin{tabular}[c]{@{}c@{}}{[}11{]}\\ $x_4\!=\!1\!-\!j$\end{tabular}                   & \begin{tabular}[c]{@{}c@{}}{[}10{]}\\ $k=3$ \end{tabular}                  & \begin{tabular}[c]{@{}c@{}}{[}01{]}\\ $\ell=2$\end{tabular}                       \\ \hline
\rowcolor[HTML]{FFFFC7} 
\begin{tabular}[c]{@{}c@{}}$M=8$\\ $m_{rf}=3$\\ $n_T=16$\end{tabular} & \begin{tabular}[c]{@{}c@{}} {[}0101010100{]} \\ $\eta=10$ \end{tabular}                                                                & \begin{tabular}[c]{@{}c@{}}{[}010{]}\\ $x_3\!=\!-1\!+\!j$\end{tabular}                   & \begin{tabular}[c]{@{}c@{}}{[}101{]}\\ $k=6$ \end{tabular}                  & \begin{tabular}[c]{@{}c@{}}{[}0100{]}\\ $\ell=5$\end{tabular}                      \\ \hline \hline
\end{tabular}
\end{table}

\subsection{RIS-SMBM Signal Model}

Fig. \ref{system-model} shows the system model of the RIS-SMBM system, which consists of one source terminal (ST), one RIS, and one destination terminal (DT). Due to an obstacle, the ST and the DT communicate with each other through the RIS. We consider the RIS to be connected to a microcontroller through a communication-oriented software and consists of $N$ reconfigurable intelligent reflecting elements which are deployed to assist data transmission to the DT by reflecting an incident RF signal transmitted by an ST. In the proposed system, ST has $n_T$ transmit  antennas, and the RT adopts single-antenna reception for a simple receiver structure. Each transmit antenna is equipped with $m_{rf}$ RF mirrors. In  the RIS-SMBM system,  the on/off status of the available  $m_{rf}$ RF mirrors are determined by the incoming $m_{rf}$ bits. Since each mirror can be on or off, a total of $2^{m_{rf}}$ different mirror activation patterns (MAPs) are occurred. Thus, different channel realizations up to  $\mathcal{F} \triangleq 2^{m{_{rf}}}$, which is also the size of  the MBM channel alphabet, can be occurred for each transmit antenna. In Fig. \ref{system-model}, $h^{n}_{\ell,k}$ and $g_{n}$  are the fading channels between the ST and the RIS, and between the RIS and DT for the $\ell$th transmit antenna, $k$th MAP state and $n$th reflecting meta-surface respectively, where $\ell \in \{1, 2, \dots n_T\}$, $k \in \{1,2, \dots \mathcal{F}\}$, and $n \in \{1, 2, \dots N\}$. For the $\ell$th transmit antenna and $k$th MAP state, channel vectors between the ST$\to$RIS, and the RIS$\to$DT can be expressed as $\mathbf{h}_{\ell,k}=\big[h^{1}_{\ell,k},h^{2}_{\ell,k}, \dots, h^{N}_{\ell,k}\big]^T \in \mathbb{C}^{ N \times 1}$ and $\mathbf{g}=\big[g_{1},g_{2}, \dots, g_{N}\big]^T \in \mathbb{C}^{N \times 1}$. We have $h^{n}_{\ell,k}$, $g_{n} \thicksim \mathcal{CN}(0, 1)$, and here $\mathcal{CN}(0, \sigma^2)$ represents  complex Gaussian distribution has zero mean and $\sigma^2$ variance. It is supposed to have perfect channel state information in DT.


%


\setcounter{equation}{5}
\begin{figure*}
\begin{eqnarray}\label{eq5}
 \mathbb{H} = \bigg[ \underset{k=1\textit{th} \ \text{channel state}}{{\underleftrightarrow{\overset{\ell =1}{\overbrace{ \textbf{h}^T_{1,1} \boldsymbol{\Phi}  \textbf{g}}}, \overset{\ell =2}{\overbrace{ \textbf{h}^T_{1,2} \boldsymbol{\Phi}  \textbf{g}}},\ldots,\overset{\ell =n_T}{\overbrace{\textbf{h}^T_{1,n_T} \boldsymbol{\Phi}  \textbf{g}}}}}} \text{ \textbrokenbar} \ \underset{k=2\textit{th} \ \text{channel state}}{{\underleftrightarrow{\overset{\ell =1}{\overbrace{ \textbf{h}^T_{2,1} \boldsymbol{\Phi}  \textbf{g}}}, \overset{\ell =2}{\overbrace{ \textbf{h}^T_{2,2} \boldsymbol{\Phi}  \textbf{g}}}, \ldots,\overset{\ell =n_T}{\overbrace{\textbf{h}^T_{2,n_T} \boldsymbol{\Phi}  \textbf{g}}}}}}\text{ \textbrokenbar} \ \ldots \ \text{ \textbrokenbar}  \ \underset{k= \mathcal{F} \textit{th} \ \text{channel state}}{{\underleftrightarrow{\overset{\ell =1}{\overbrace{ \textbf{h}^T_{\mathcal{F},1} \boldsymbol{\Phi}  \textbf{g}}},\overset{\ell =2}{\overbrace{ \textbf{h}^T_{\mathcal{F},2} \boldsymbol{\Phi}  \textbf{g}}},\ldots,\overset{\ell =n_T}{\overbrace{\textbf{h}^T_{\mathcal{F},n_T} \boldsymbol{\Phi}  \textbf{g}}}}}} \bigg]_{1 \times \mathcal{F} n_T}
 \\  \hline \nonumber
\end{eqnarray} 
\end{figure*}
\setcounter{equation}{0}

The RIS-SMBM scheme makes use of the MBM technique to increase the spectral efficiency of the SM scheme. Therefore, the RIS-SMBM system utilizes two indices which are the active transmit antenna index and the active MAP index to convey information bits stream. In Fig. \ref{system-model}, the incoming data bits expressed by the vector $\textbf{b}$ have dimensions $1 \times \eta$ is the sequence of data bits to be conveyed during one symbol period ($T_s$). In the proposed system, $\textbf{b}$ is split into three subgroups, which consist of the following there parts: 
\begin{itemize}
	\item $m_\text{S} = {\log}_2(M)$ bits map to a $M$-QAM symbol $x_p$, here $p \in \{1, 2, \dots M\}$, 
	\item $m_{rf}$ bits are used to select the $k$th MAP,
	\item and the remaining $m_{\text{SM}}= \log_2(n_T)$ bits are mapped to select a transmit antenna with index $\ell$.
\end{itemize}
For these subgroups, $m_\text{S}$ bits are transmitted an $M$-QAM symbol, whereas $m_{rf}$ bits are conveyed with MAPs state, and the $m_{\text{SM}}$ bits are transmitted in the active antenna index of the SM system. Finally, as can be shown from the transmitter of the RIS-SMBM scheme, the selected symbol is conveyed over the MAP state, and the single transmit antenna is activated by the SM technique. 

Eventually, the spectral efficiency of the RIS-SMBM scheme can be written as follows:
\begin{eqnarray}\label{eq1}
\eta ={{\mathrm{log}}_2 \left(M\right)\ }+m_{rf}+ {{\mathrm{log}}_2 \left(n_T\right) \ } \text{[bits/s/Hz]}.
\end{eqnarray}

A tabular representation of the bit mapping for the RIS-SMBM system is presented in Table \ref{T1}. Considering Table \ref{T1}, an example of the mapping procedure of RIS-SMBM can be given. Suppose that the  bit stream  $\textbf{b}= \big[111001\big]$ will be transmitted by the ST using the following system parameters: $M=4,m_{rf}=4,n_T=4$ and $\eta=8$ bits. Assume that the data bits stream is grouped as follows:
\begin{eqnarray}\label{eq2}
\textbf{b} &=& \big[\!  \underbrace{1\ 1}_{m_\text{S}  \ \text{bits}} \  \underbrace{ \ 1 \ 0}_{m_\text{rf} \  \text{bits}} \  \underbrace{0 \ 1}_{m_\text{SM} \  \text{bits}} \!\!\big].
\end{eqnarray}
The first $\log_2(M)=2$ bits $([1\ 1])$ determine the 4-QAM symbol $x_4=1-j$. Then, next $m_{rf}$ bits ([  1 \ 0 ]) select the $k=3$ channel state that corresponds to the MAPs status of the RF mirrors for the activated transmit antenna of ST. Finally, the last $m_{\text{SM}}$ bits $([0 \ 1])$ map to the active antenna index $\ell=2$ over which the selected $x_4$ will be conveyed to DT through the RIS.

\subsection{RIS-SMBM Channel Model}
 
The received signal transmitted $\ell$th transmit antenna, $k$th MAP state and reflected through the RIS that has $N$ reflecting surfaces is expressed as follows:
\begin{eqnarray}\label{eq3}
 y &=& \Bigg[\sqrt{E_s}\sum^{N}_{n=1}h^{n}_{\ell,k} e^{j{\phi^{n}_{\ell,k}}}g_n\Bigg]\ x_p\ + w \nonumber  \\
 &=& \sqrt{E_s}\ \textbf{h}^{T}_{\ell,k} \boldsymbol{\Phi} \ \textbf{g}\ x_p + w,
\end{eqnarray}
where,  $y \in \mathbb{C}^{1 \times 1}$ is the received signal, and  $E_s$ represents the transmitted symbol energy. We have $h^{n}_{\ell,k}={\alpha }^{n}_{\ell,k}e^{-j{\theta^{n}_{\ell,k} }}$  and $g_n = {\beta }_n e^{-j{{\varphi}}_n}$. $\phi^{n}_{\ell,k}$ is the adjustable phase generated by the $n$th element of the RIS. Using the activated $\ell$th transmit antenna and $k$th MAP state, the reflector phases $\boldsymbol{\phi}_{\ell,k} = \big\{ \phi^{n}_{\ell,k} \big\}_{n=1}^N$ are adjusted according to the data bits to maximize the instantaneous SNR in the receiver. To accomplish this, After the incoming $m_\text{SM}$ and $m_{rf}$ bits specify the index $\ell$ of a transmit antenna and the index $k$ of a MAP state are determined, the RIS adjusts its phases according to this selected transmit antenna and $k$th MAP state as $\phi^{n}_{\ell,k}=({\theta }^{n}_{\ell,k}+{\varphi}_n) $ for $n \in \{1, 2, \ldots , N\}$. $\boldsymbol{\Phi}$ is a reflection matrix that contains reflection phases of the reflecting surfaces. $\boldsymbol{\Phi}$ is controlled by the RIS control signal from the ST and then, $\boldsymbol{\Phi}=\mathtt{diag} \big\{ e^{j{{\phi}^{1}_{\ell,k}}},e^{j{{\phi}^{2}_{\ell,k}}},\cdots,e^{j{{\phi}^{N}_{\ell,k}}} \big\}$. Also,  the additive white Gaussian noise is expressed as $w\thicksim\mathcal{CN}\big(0,N_0 \big)$. The received baseband signal model of the RIS-SMBM   can be also given in vector form as follows:
\begin{eqnarray}\label{eq4}
 y =\sqrt{E_s} \mathbb{H} \ \mathbf{x}\ + w,
\end{eqnarray}
where $\mathbb{H}\in \mathbb{C}^{1\times \mathcal{F}n_T}$  and $\mathbf{x}\in \mathbb{C}^{ \mathcal{F}n_T \times 1}$  express the channel matrix and transmission vector of the RIS-SMBM system, and their dimensions are $1 \times \mathcal{F} n_T$ and  $ \mathcal{F}n_T \times 1$. The extended structure of $\mathbb{H}$ is expressed in (\ref{eq5}).  $\textbf{x}$ is can be given as:
\begin{eqnarray}\label{eq6}
\textbf{x}_{\ell,k,p}  \triangleq \bigg[ 0 \cdots 0\text{ \textbrokenbar}  \cdots  \text{ \textbrokenbar}   \underset{k\text{th} \ \text{channel state}}{{\underleftrightarrow{0 \cdots 0 \! \!\! \!\! \! \overset{\ell \text{th \ position}}{\overarrow{x_p}} \! \!\! \!\! \! \!  0 \cdots 0}}}\text{ \textbrokenbar}  \cdots  \text{ \textbrokenbar}  0 \cdots 0 \bigg]^T, 
\end{eqnarray}
Also, note that depending on the $\ell$ and $k$ parameters, the position (index) of $x_p$ in \textbf{x} can be written as:
\setcounter{equation}{6}
\begin{eqnarray}\label{eq_i}
i = \big( k-1 \big)n_T \ + \ \ell, 
\end{eqnarray}
For the above example, the corresponding transmission vector is expressed as follows:
 \begin{eqnarray}\label{eq7}
\textbf{x}_{\ell,p} &=& \bigg[ \underset{1\text{th} \ \text{ch. state}}{{\underleftrightarrow{0\ 0\ 0\ 0}}}\text{ \textbrokenbar} \underset{2\text{th} \ \text{ch. state}}{{\underleftrightarrow{0\ 0\ 0\ 0}}}  \text{ \textbrokenbar}   \underset{3\text{th} \ \text{ch. state}}{{\underleftrightarrow{0\ \! \!\! \!\! \!\! \! \overset{\ell =2 \text{th \ position}}{\overarrow{1-j}} \! \!\! \!\! \!\! \!\ 0\ 0}}}\text{ \textbrokenbar}    \underset{4\text{th} \ \text{ch. state}}{{\underleftrightarrow{0\ 0\ 0\ 0}}} \bigg]^T\!\!\!. 
\end{eqnarray}
Using equation (7) for the above example, the position of $x_p$ can be calculated as $i = \big( k-1 \big)n_T \ + \ \ell = \big( 3-1 \big)4 \ + \ 2 = 10$.

The instantaneous SNR expression of the RIS-SMBM system is defined as follows:
\begin{eqnarray} \label{SNR} 
\gamma  &=& \frac{ {\sqrt{E_s} \ \bigg|\sum\limits_{n=1}^{N}\alpha^n_{\ell,k}\beta_{n} e^{(j\phi^n_{\ell,k}-j\theta^{n}_{\ell,k}-j{\varphi}_n)}\bigg|}^2}{N_0},
\end{eqnarray} 

To maximize the instantaneous SNR value in the equation in (\ref{SNR}), the adjustable phase $\phi^n_{\ell,k}$ should be selected to zero the $\theta^{n}_{\ell,k}$ and $\varphi_n$ channel phases. Therefore, RIS in Fig. \ref{system-model} selects the adjustable phase value as $\phi^n_{\ell,k}=\theta^{n}_{\ell,k}+\varphi_n$. As a result, the transmitted signal quality is increased by maximizing the instantaneous SNR value thanks to RIS.

\subsection{ML Detector for RIS-SMBM System}

With $\textbf{h}_{\ell,k}$, $ \boldsymbol{\Phi}$ and $ \textbf{g}$ known at the receiver, the ML detector for the RIS-SMBM scheme is written as follows:
\begin{eqnarray}\label{eq8}
 \Big [\hat{x},\hat{\ell},\hat{k}\Big] &=& \arg\max_{(p,\ell,k)}  P_y\Big( y \ \big| \ \textbf{h}_{\ell,k}, \boldsymbol{\Phi}, \ \textbf{g},\ x_p \Big)\nonumber  \\
 &=& \arg\max_{(\ell,k,p)}  P_y\Big( y \ \big| \ \mathbb{H}, \ \mathbf{x}\Big) \nonumber  \\
  &=& \arg\min_{(\ell,k,p)}  \Big|\Big| y- \sqrt{E_s}\ \textbf{h}^{T}_{\ell,k} \boldsymbol{\Phi} \ \textbf{g}\ x_p\Big|\Big|^2 \nonumber  \\
    &=& \arg\min_{(\ell,k,p)} \ \Big|\Big| y- \sqrt{E_s} \mathbb{H} \ \mathbf{x} \ \Big|\Big|^2,
\end{eqnarray}
since $P_y\big( y  \big|  \textbf{h}_{\ell,k}, \boldsymbol{\Phi}, \ \textbf{g}, x_p \big) \propto \exp{\Big(-\big|\big| y- \sqrt{E_s}\ \textbf{h}^{T}_{\ell,k} \boldsymbol{\Phi}  \textbf{g} x_p\big|\big|^2\Big)}$.

After estimating the $p$, $\ell$, and $k$ parameters (i.e., $\hat{p},\hat{\ell},\hat{k}$), the estimated bit sequence $\hat{\textbf{b}}$ by demapper $\&$ sort and merge function is generated at the destination side, as seen in Fig. \ref{system-model}.

\subsection{ELC Detector for RIS-SMBM System}

ELC detector estimates the active antenna index, the active MAP index, and the transmitted symbol over the received signal with a lower complexity cost than the optimal ML detector. The ELC detector of the RIS-SMBM system is defined as follows \cite{enhanced}:
\begin{eqnarray}\label{eq_enhanced}
\!\!\!\!\!\!\!\!\! \Big[\hat{x},\hat{\ell},\hat{k}\Big] & \!\!\!\!=& \!\!\!\! \text{arg}\underset{p,\ell,k}{\mathrm{max}} \Bigg\{2\Re\Big(\mathbf{h}_{\ell,k}^\text{H} \boldsymbol{\Phi}  \textbf{g}yx_p^*\Big) - \big| x_p \big|^2 \big| \mathbf{h}_{\ell,k}^T \boldsymbol{\Phi} \textbf{g} \big|^2 \Bigg\}, 
\end{eqnarray}
where, $x_p^*$ is the conjugate of $x_p$ complex symbol. After determining the $x$, $l$, and $k$ values with lower complexity detector, the estimated bit sequence $\hat{\textbf{b}}$ is generated by using demapper $\&$ sort and merge block as seen in Fig. \ref{system-model}.





\section{Performance Analysis of RIS-SMBM System}


In this section, the ABER of the RIS-SMBM technique is analyzed.  The original bits of information transmitted from the transmitter may be obtained erroneously at the receiver side, due to the fading effect of the Rayleigh channel and the noise. Examining the pairwise error probability (PEP), which gives information about the probability of making an erroneous decision, is important for analytical error analysis. By using the well-known upper bounding technique, the analytical ABER of the RIS-SMBM system is described as follows:
\begin{equation} \label{ABER} 
	\mathbb{P}_{\text{RIS-SMBM}} \approx \frac{1}{\eta2^\eta}\sum^{2^\eta}_{i=1}{\sum^{2^\eta}_{j=1}{\mathbb{P}\big( \textbf{x}_i \rightarrow \hat{\textbf{x}}_j \big)\mspace{2mu}e_{i,j}}}, 
	\end{equation} 
where, $\mathbb{P}\big( \textbf{x}_i \rightarrow \hat{\textbf{x}}_j \big)$ represents the average PEP given in (\ref{ABER}), and $e_{i,j}$ is defined as the number of bit errors associated with the corresponding pairwise error event. 

When $\textbf{x}_{i,p}$ is transmitted and it is erroneously detected as $\textbf{x}_{\hat{i},\hat{p}}$, the conditional PEP (CPEP) of the proposed system is given as follows:
\begin{eqnarray}
 \mathbb{P}\Big( \textbf{x}_{i,p} \rightarrow \ \textbf{x}_{\hat{i},\hat{p}} \ \big| \ \mathbb{H} \Big)  &=&  \mathbb{P}\Bigg(\big\vert y- \mathbb{H}  \textbf{x}_{i,p}\big\vert^2 > \big\vert y- \mathbb{H} \textbf{x}_{\hat{i},\hat{p}}  \big\vert^2\Bigg)      \nonumber\\
 &\!\!\!\!\!\!\!\!\!\!\!\!\!\!\!\!\!\!\!\!\!\!\!\!\!\!\!\!\!\!\!\!\!\!\!\!\!\!\!\!\!\!\!\!\!\!\!\!\!\!\!\!\!\!\!\!\!\!\!\!\!\!\!\!\!\!\!\!\!\!\!\!\!\!\!\!\!\!\!\!\!\!\!\!\!\!\!\!\!\!\!\!\!\!\!\!\!\!\!\!\!\!\!\!\!\!\!\!\!=&\!\!\!\!\!\!\!\!\!\!\!\!\!\!\!\!\!\!\!\!\!\!\!\!\!\!\!\!\!\! \!\!\!\!\!\!\!\!\!\!\!\!\!\!\!\!\!\!\!\!\!\!\!\!\!\!\!\mathbb{P}\Bigg(\big\vert {w}\big\vert^2  - \Big(\mathbb{H} \big( \textbf{x}_{i,p} -  \textbf{x}_{\hat{i},\hat{p}} \big) + {w} \Big)  
    \Big( \mathbb{H} \big( \textbf{x}_{i,p} - \textbf{x}_{\hat{i},\hat{p}} \big)  + {w} \Big)^H > 0 \Bigg) \nonumber\\
   &\!\!\!\!\!\!\!\!\!\!\!\!\!\!\!\!\!\!\!\!\!\!\!\!\!\!\!\!\!\!\!\!\!\!\!\!\!\!\!\!\!\!\!\!\!\!\!\!\!\!\!\!\!\!\!\!\!\!\!\!\!\!\!\!\!\!\!\!\!\!\!\!\!\!\!\!\!\!\!\!\!\!\!\!\!\!\!\!\!\!\!\!\!\!\!\!\!\!\!\!\!\!\!\!\!\!\!\!\!=&\!\!\!\!\!\!\!\!\!\!\!\!\!\!\!\!\!\!\!\!\!\!\!\!\!\!\!\!\!\! \!\!\!\!\!\!\!\!\!\!\!\!\!\!\!\!\!\!\!\!\!\!\!\!\!\!\! \mathbb{P}\Bigg( \big| \mathbb{H}\big( \textbf{x}_{i,p} - \textbf{x}_{\hat{i},\hat{p}}\big) \big|^2  - 2\Re \Big\{ {w}^H  \Big(\mathbb{H} \big( \textbf{x}_{i,p} - \textbf{x}_{\hat{i},\hat{p}} \big) \Big) \Big\} > 0 \Bigg) \nonumber  \\
    &\!\!\!\!\!\!\!\!\!\!\!\!\!\!\!\!\!\!\!\!\!\!\!\!\!\!\!\!\!\!\!\!\!\!\!\!\!\!\!\!\!\!\!\!\!\!\!\!\!\!\!\!\!\!\!\!\!\!\!\!\!\!\!\!\!\!\!\!\!\!\!\!\!\!\!\!\!\!\!\!\!\!\!\!\!\!\!\!\!\!\!\!\!\!\!\!\!\!\!\!\!\!\!\!\!\!\!\!\!=&\!\!\!\!\!\!\!\!\!\!\!\!\!\!\!\!\!\!\!\!\!\!\!\!\!\!\!\!\!\! \!\!\!\!\!\!\!\!\!\!\!\!\!\!\!\!\!\!\!\!\!\!\!\!\!\!\!\mathbb{P} \Big( \mathcal{U} > 0 \Big),
\label{eq:qEquation_ea}
\end{eqnarray}
where $\mathcal{U}\sim \mathcal{N}(\mu_\mathcal{U},\sigma^2_\mathcal{U})$ is a Gaussian distributed R.V, where $\mu_\mathcal{U}$ and $\sigma^2_\mathcal{U}$ can be  given as follows respectively:
\begin{eqnarray}
   \mu_\mathcal{U}&=&  E\Bigg\{\big| \mathbb{H}\big( \textbf{x}_{i,p} - \textbf{x}_{\hat{i},\hat{p}}\big) \big|^2  - 2\Re \Big\{ {w}^H  \Big(\mathbb{H} \big( \textbf{x}_{i,p} - \textbf{x}_{\hat{i},\hat{p}} \big) \Big) \Big\} \Bigg\} \nonumber \\ 
 &=&  E\Big\{\big| \mathbb{H}\big( \textbf{x}_{i,p} - \textbf{x}_{\hat{i},\hat{p}}\big) \big|^2  \Big\}  \nonumber \\ 
  &=&  \big| \mathbb{H}\big( \textbf{x}_{i,p} - \textbf{x}_{\hat{i},\hat{p}}\big) \big|^2 ,
\end{eqnarray}
\begin{eqnarray}
  \sigma^2_{\mathcal{U}}  &=&  Var\Big\{\big| \mathbb{H}\big( \textbf{x}_{i,p} - \textbf{x}_{\hat{i},\hat{p}}\big) \big|^2  - 2\Re \Big\{ {w}^H  \Big(\mathbb{H} \big( \textbf{x}_{i,p} - \textbf{x}_{\hat{i},\hat{p}} \big) \Big) \Big\}\Big\}    \nonumber \\
  &=&  4 \big| \mathbb{H}\big( \textbf{x}_{i,p} - \textbf{x}_{\hat{i},\hat{p}}\big) \big|^2Var\big( w^H \big) \nonumber\\
 &=& 2N_0 \big| \mathbb{H}\big( \textbf{x}_{i,p} - \textbf{x}_{\hat{i},\hat{p}}\big) \big|^2.
\end{eqnarray}

Considering the proposed RIS-SMBM system model, depending on the correct and incorrect detection of the index $i$ in (\ref{eq_i}), which includes the information of active MAP and active antenna indices, two cases occur.


\emph{Case 1:} Incorrect detection of the index $i$, $i\neq\hat{i}$ \newline
In the first case, when the index $i$ is detected incorrectly, the CPEP function can be written in terms of $Q(.)$ as follows:
\begin{equation}\label{eq_cpep_1}
\begin{aligned}
\mathbb{P}\Big( \textbf{x}_{i,p} \rightarrow \ \textbf{x}_{\hat{i},\hat{p}} \ \big| \ \mathbb{H} \Big) &=& Q\Bigg(  \sqrt{\frac{\big| \mathbb{H} \big( {\textbf{x}_{i,p}} - \textbf{x}_{\hat{i},\hat{p}} \big) \big|^2}{2N_0}}  \Bigg).   
\end{aligned}
\end{equation}

\emph{Case 2:} Correct detection of the index $i$ , $i=\hat{i}$ \newline
In the second case, when the index $i$ is correctly estimated, the CPEP function is expressed as follows:
\begin{equation}\label{eq_cpep_2}
\begin{aligned}
\mathbb{P}\Big( \textbf{x}_{i,p} \rightarrow \ \textbf{x}_{\hat{i},\hat{p}} \ \big| \ \mathbb{H} \Big)  &=& Q\Bigg(  \sqrt{\frac{\big| \mathbb{H} \big( {\textbf{x}_{i,p}} - \textbf{x}_{i,\hat{p}} \big)  \big|^2}{2N_0}} \Bigg).   
\end{aligned}
\end{equation}

Unconditional PEP (UPEP) for case 2 (it can also be simply written for case 1) can be given as follows:
\begin{eqnarray}\label{eq_pep}
\mathbb{P}\Big( \textbf{x}_{i,p} \rightarrow \ \textbf{x}_{\hat{i},\hat{p}} \ \Big)  &=& E_{ \mathbb{H}}\Bigg\{ \mathbb{P}\Big( \textbf{x}_{i,p} \rightarrow \ \textbf{x}_{\hat{i},\hat{p}} \ \big| \ \mathbb{H} \Big)  \Bigg\} \nonumber \\
&=& E_{ \mathbb{H}}\Bigg\{Q\Bigg(  \sqrt{\frac{\big| \mathbb{H} \big( {\textbf{x}_{i,p}} - \textbf{x}_{i,\hat{p}} \big)  \big|^2}{2N_0}} \Bigg)\Bigg\},   
\end{eqnarray}
where $E_{ \mathbb{H}}\{.\}$ is the expectation with respect to the channel vector ${ \mathbb{H}}$.

Finally, the analytical ABER of the RIS-SMBM system is described as follows:
\begin{eqnarray} \label{ABER_v1} 
\hspace{-0.4cm}\mathbb{P}_{\text{RIS-SMBM}} &\!\!\!\!\approx &\!\!\!\!\! \frac{1}{\eta2^\eta}\!\!\sum^{2^\eta}_{i=1}{\sum^{2^\eta}_{j=1}{\!\!E_{ \mathbb{H}}\Bigg\{\!Q\Bigg( \! \sqrt{\frac{\big| \mathbb{H} \big( {\textbf{x}_{i,p}} - \textbf{x}_{j,\hat{p}} \big)  \big|^2}{2N_0}} \Bigg)\!\Bigg\}e_{i,j}}}.
	\end{eqnarray}

\definecolor{aureolin}{rgb}{0.6, 0.99, 0.99}
\definecolor{arylideyellow}{rgb}{0.85, 1,0.99 }

\begin{table}[t]

\addtolength{\tabcolsep}{3.5pt}
		\caption{Energy-saving comparisons of RIS-SMBM system in percentage compared to RIS-SM, RIS-MBM, and RIS-QSM systems ($\mathcal{E}_{\text{sav}}$ \%).}
		\vspace{-1em}
	\begin{center}
		\label{Tabloenergy}
		\begin{tabular}{|ccc||ccc|} 
		\rowcolor{aureolin}
		    \hline
	        \hline
			
			$M$ & $n_T$ & $m_{rf}$   & RIS-SM & RIS-MBM & RIS-QSM \\ 
		\rowcolor{arylideyellow}
			\hline
			\hline
			$8$ & $4$ & $5$   & $50.00\%$ & $20.00\%$ & $30.00\%$ \\  
		\rowcolor{arylideyellow}
			$16$ & $16$ & $10$   & $55.56\%$   & $22.22\%$ & $33.33\%$ \\   
		\rowcolor{arylideyellow}
			$32$ & $64$ & $15$  & $57.69\%$ & $23.08\%$ & $34.62\%$ \\   
			\hline
			\hline
			
		\end{tabular}
	\end{center}
		\vspace{-1em}
\end{table}

\section{The Energy Efficiency, Throughput, Data Rate, and Complexity Analysis for RIS-SMBM System}

In this section, energy efficiency, throughput, data rate, and complexity analyzes, which are important performance criteria, are obtained. The energy efficiency, throughput, data rate, and complexity analyzes of the RIS-SMBM scheme are compared with the RIS-SM, RIS-MBM, and RIS-QSM systems.

\subsection{The Energy Efficiency Analysis}	
	
In the proposed RIS-SMBM system, most of the information bits are carried in the active transmit antenna index and the active MAP index. When information bits are transmitted in indices, little or no energy is consumed. Therefore, the RIS-SMBM system provides high energy efficiency. The percentage of energy-saving $(\mathcal{E}_{\text{sav}})$ per $\eta$ bits of the RIS-SMBM system compared to other benchmark systems is defined as follows:
\begin{equation}\label{energy}
\mathcal{E}_{\text{sav}}=\Big(1-\frac{n_b}{\eta}\Big)E_b\%,
\end{equation}
where $E_b$ is the bit energy and $n_b$ is the spectral efficiency of benchmark systems. The percentage of energy efficiency provided by the RIS-SMBM system compared to the RIS-SM, RIS-MBM, and RIS-QSM benchmark systems is presented in Table \ref{Tabloenergy}. Three different cases are considered for different $M$, $n_T$, and $m_{rf}$ values, and the percentage of energy efficiency provided by the RIS-SMBM system is calculated for all cases. It is seen that the RIS-SMBM system provides higher energy efficiency than all the benchmark systems. For example, for the first case $M=8$, $n_T=4$ and $m_{rf}=5$ in Table \ref{Tabloenergy}, the RIS-SMBM system provides $50\%$, $20\%$, and $30\%$ higher energy efficiency than RIS-SM, RIS-MBM, and RIS-QSM systems, respectively.   

\begin{table}[t]
\caption{Data Rate Comparisons in [bits/s/Hz] for RIS-SM, RIS-MBM, RIS-QSM, and RIS-SMBM schemes}
\begin{center}
	\label{data_rate}
 \begin{tabular}{|ccc||cccc|}
\rowcolor{amber}
\hline
\hline

$n_T$ &  $m_{rf}$ & $M$ & RIS-SM & RIS-MBM & RIS-QSM &  RIS-SMBM \\
\rowcolor{blond}
\hline
\hline   
$2$  &  $8$      & $16$      & $5$           & $12$  &  $6$ &  $13$        \\
\rowcolor{blond}
\hline
$8$  &  $5$        & $4$       &    $5$  &  $7$  &  $8$ & $10$   \\
\rowcolor{blond}
\hline 
$32$  &  $10$  & $8$ &  $9$ &  $13$ & $13$ &  $18$   \\ 
\rowcolor{blond}
\hline
\hline
\end{tabular}
\end{center}
\vspace{-1em}
\end{table}

\definecolor{aureolin}{rgb}{0.7,1 , 0.6}
\definecolor{arylideyellow}{rgb}{0.9,1 , 0.8}

\begin{table}[t]
		\caption{A Comparative Evaluation of the Computational Complexity}
	\begin{center}
	\addtolength{\tabcolsep}{17pt}
		\label{Tablocomplexity}
		\begin{tabular}{|c||c|} 
			\rowcolor{aureolin}
		 
	        \hline
			\hline
			  Systems & Real Multiplications (RMs) \\ 	
		
			\rowcolor{arylideyellow}
			\hline
			\hline
			
			\!\!$\mathcal{O}_{\text{ RIS-SMBM (ML)}}$\!\!  &  ${(N+4M)n_T2^{m_{rf}}}$  \\   
			\rowcolor{arylideyellow}		
				\!\!$\mathcal{O}_{\text{ RIS-SMBM (ELC)}}$\!\!  &  ${3(1+\frac{(M+N)}{4})n_T2^{m_{rf}}}$  \\   
			\rowcolor{arylideyellow}	
				
			$\mathcal{O}_{\text{ RIS-SM}}$  &  $(N+4M){n_T}\bigg(1+\dfrac{m_{rf}}{n_S+n_{SM}}\bigg)$ \\  
			\rowcolor{arylideyellow}
			$\mathcal{O}_{\text{ RIS-QSM}}$   &  $(N+4M){n_T}\bigg(1+\dfrac{m_{rf}}{n_S+2n_{SM}}\bigg)$  \\ 
			\rowcolor{arylideyellow}						$\mathcal{O}_{\text{ RIS-MBM}}$  &  ${(N+4M)2^{m_{rf}}}\bigg(1+\dfrac{n_{SM}}{n_S+m_{rf}}\bigg)$  \\ 
			\hline
			\hline
	
		\end{tabular}
	\end{center}
\end{table}

\begin{figure}[t]
\centering
\includegraphics[scale=0.6]{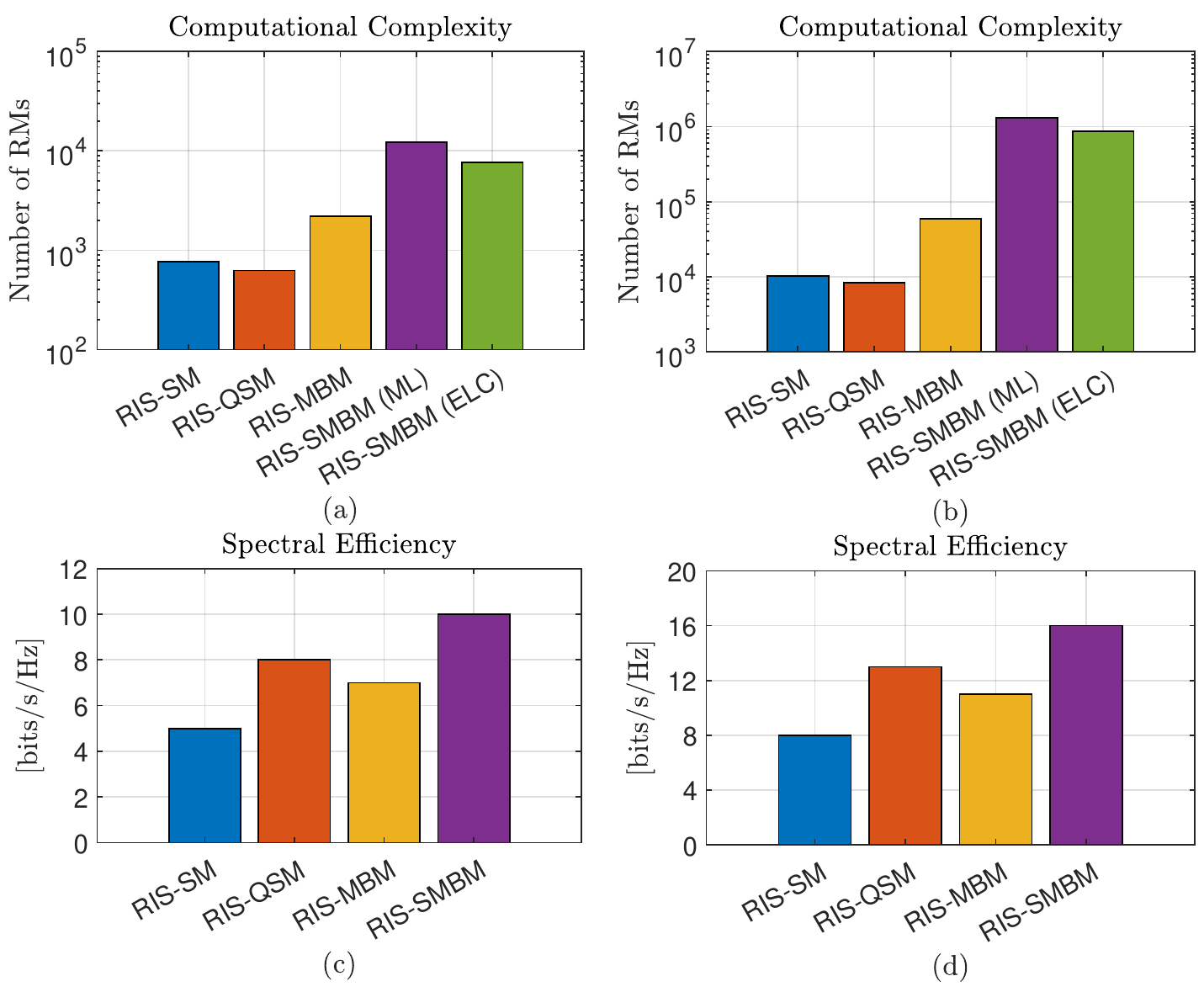}
 \caption{Computational Complexity and Spectral Efficiency Comparisons while $M=4$, $N_T=8$, $m_{rf}=5$, $N=32$  for (a), (c); and $M=8$, $N_T=32$, $m_{rf}=8$, $N=128$   for (b), (d).}
\label{complx_se}
\end{figure}

\subsection{The Throughput and Data Rate Analysis}

The throughput for the considered system is defined as the number of bits that can be correctly obtained at the receiver from the bits. According to this definition, the throughput for the RIS-SMBM system is expressed as \cite{Tse}:
\begin{equation}\label{throughput}
\mathcal{T}    =  \frac{\big(1-\mathbb{P}_{\text{RIS-SMBM}}\big)}{T_s} \eta,
\end{equation}
where $(1-\mathbb{P}_{\text{RIS-SMBM}})$ is defined as the probability of correct bits obtained by the receiver of the RIS-SMBM system during the symbol transmission period $T_s$. Considering Table \ref{data_rate}, it is seen that the amount of information bits transmitted during a symbol period of the RIS-SMBM system is higher than the RIS-SM, RIS-MBM, and RIS-QSM systems. Therefore, it is clear that the proposed RIS-SMBM system provides high throughput. 



\subsection{Complexity Analysis}

This section offers the receiver computational complexities of the proposed RIS-SMBM system and benchmark systems. Computational complexity expressions of RIS-SMBM, RIS-SM, RIS-QSM, and RIS-MBM systems are given in Table \ref{Tablocomplexity}. The computational complexity of the systems is calculated in terms of real multiplications (RMs). The computational complexity of the proposed RIS-SMBM system is higher than benchmark systems, since both the transmit antenna index and the MAP index are used in information transmission. However, despite its high complexity, the RIS-SMBM system provides higher spectral efficiency and better performance than benchmark systems. Also, numerical examples of spectral efficiency and computational complexity of RIS-SM, RIS-QSM, RIS-MBM, and RIS-SMBM systems are given in Fig. \ref{complx_se}. As can be easily seen from Fig. \ref{complx_se}, the RIS-SMBM system has higher spectral efficiency than other systems. However, the RIS-SMBM (ML) system has considerably higher computational complexity than other systems. Therefore, an ELC detector is proposed to reduce the complexity of the RIS-SMBM system. It can be seen in Fig. \ref{complx_se} that the ELC detector  reduces the complexity of the RIS-SMBM system.



 \section{Simulation Results and Discussions}

In this section, simulation results for the proposed RIS-SMBM system are presented. The proposed system is compared with its counterparts and the results are discussed. Comparison curves of the analytical and BER performances  of the proposed system are given and performance changes are investigated in varying parameters. The SNR expression in the simulations is defined as $\mathrm{SNR (dB)}=10\log_{10}(E_s/N_0)$. The optimal ML and ELC detector for the RIS-SMBM system are used for the estimation of the transmitted symbol and active indices at the receiver. All simulation results are obtained under the assumption that fading channels are uncorrelated Rayleigh distributions. Also, for all systems, the receiver antenna is selected as one.

\begin{figure}[t]
\centering{\includegraphics[width=0.476\textwidth]{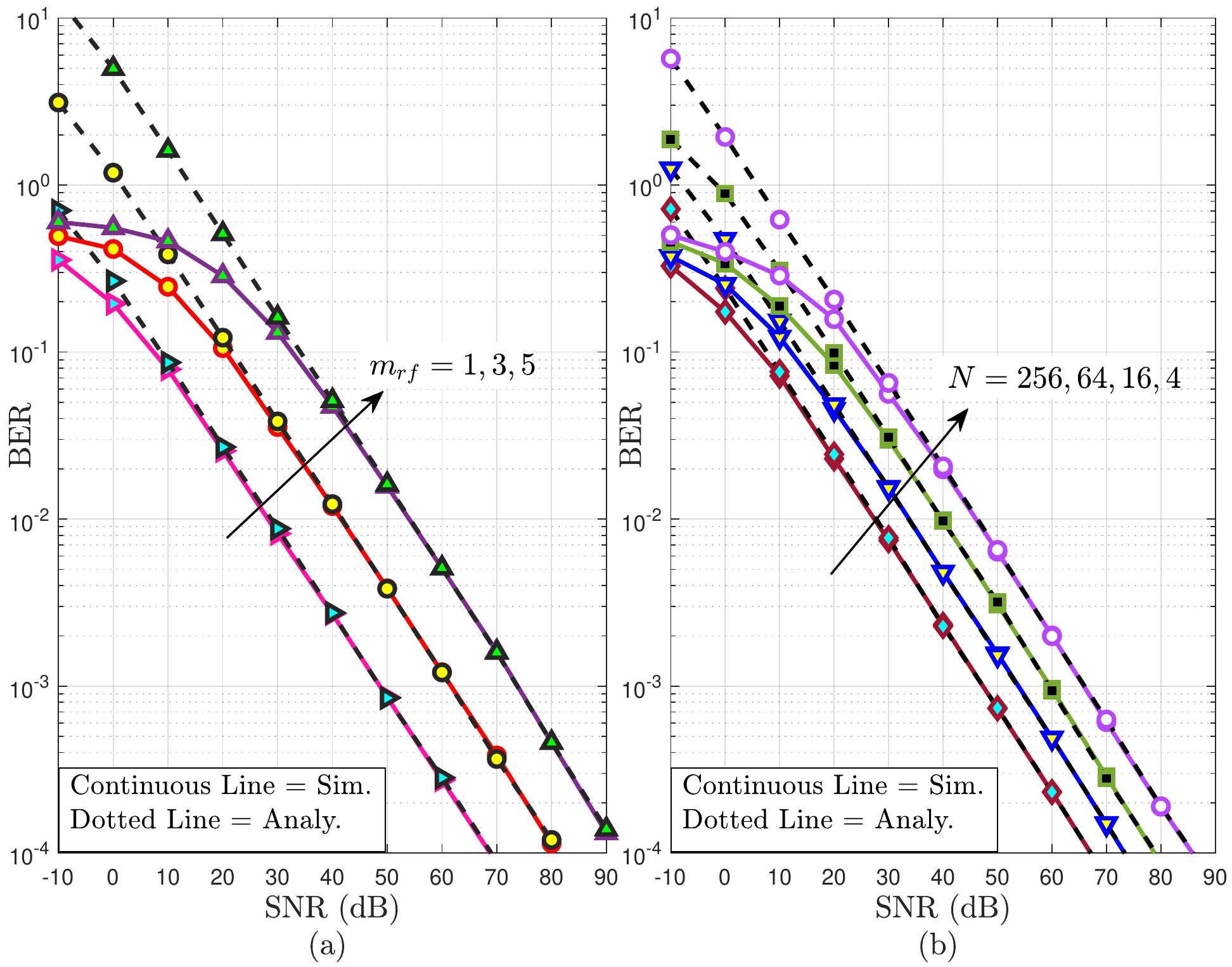}}
 \vspace{-0.675em}
	\caption{Performance comparisons of the RIS-SMBM system (a) for $M=2$, $n_T=4$, $N=64$ while $m_{rf}=1,3,5$, (b) for $M=4$, $n_T=4$, and $m_{rf}=2$ while $N=4,16,64,256$.}

	\label{SMBMRIS_RFs} 
\end{figure}

In Fig. \ref{SMBMRIS_RFs} (a), the BER performance curves of the RIS-SMBM system are presented for varying $m_{rf}$ values while $M=2$, $n_T=64$, $N=64$. The spectral efficiency of the RIS-SMBM system is $\eta=4,6,8$ while $m_{rf}=1,3,5$, respectively. When the obtained results are examined, as the $m_{rf}$ number decreases by 2, an improvement of approximately 13 dB is observed in the BER results. As can be observed from Fig. \ref{SMBMRIS_RFs} (a), when the number of $m_{rf}$ decreases, the number of transmitted information decreases, but the BER performance improves. The BER performances of the proposed RIS-SMBM system in Fig. \ref{SMBMRIS_RFs} (b) are given and compared with analytical results for $M=4$, $n_T=4$, and $m_{rf}=2$ at varying $N$ values. The simulation and analytical results are quite similar to each other. Also, as can be easily seen from the BER curves in Fig. \ref{SMBMRIS_RFs} (b), it is clear that the performance of the system improves as the $N$ parameter increases. Thus, using a reasonable number of reflecting surfaces $N$ will enable the system to function at the desired error rates. Also, in Fig. \ref{SMBMRIS_RFs} (a) and (b), the simulation and analytical curves overlap perfectly.

\begin{figure}[t]
\centering{\includegraphics[width=3.3 in]{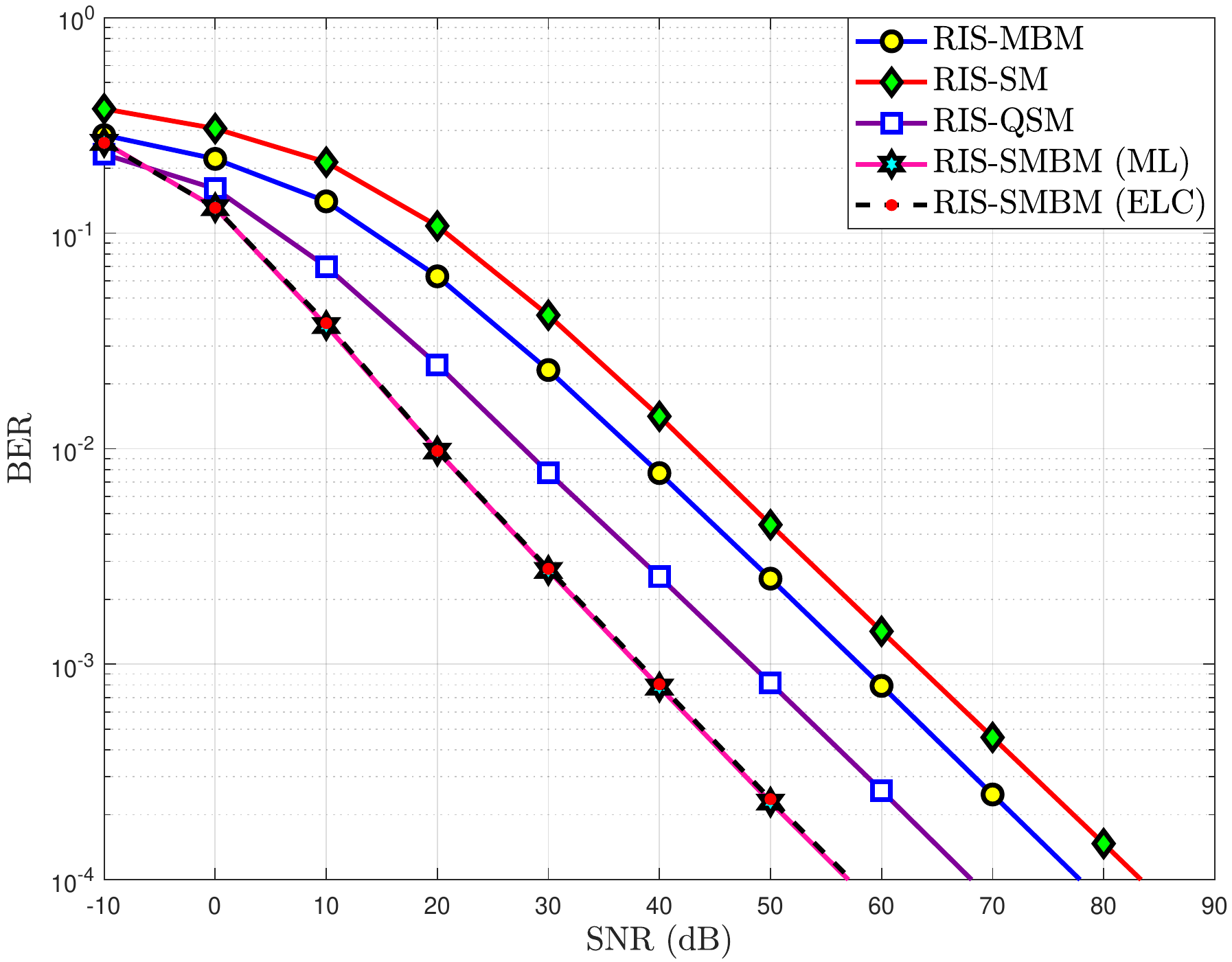}}
	\caption{Simulation performance comparisons of the RIS-SMBM, RIS-MBM, RIS-SM, and RIS-QSM systems while $N=128$ and $\eta=8$ for all systems. }
	\label{SMBMRIS_benchmarks_n8} 
\end{figure}

We present BER performance comparisons of RIS-SM, RIS-MBM, RIS-QSM, and RIS-SMBM systems for $\eta=8$ bits/s/Hz and $N=128$ in Fig. \ref{SMBMRIS_benchmarks_n8}. The system parameters selected for the RIS-SMBM, RIS-MBM, RIS-SM, and RIS-QSM systems are ($M=64$, $n_T=2$, $m_{rf}=1$), ($M=8$, $m_{rf}=5$), ($M=4$, $n_{T}=64$) and ($M=4$, $n_{T}=8$), respectively.  It is seen that the RIS-SMBM system provides better BER performance than RIS-SM, RIS-MBM, and RIS-QSM systems at the same spectral efficiency. RIS-SMBM system provides $11.10$ dB, $20.86$ dB, and $26.35$ dB SNR gain according to RIS-QSM, RIS-MBM, and RIS-SM systems, respectively. It is also seen that the BER performances of the RIS-SMBM (ML) and RIS-SMBM (ELC) systems are almost the same.

\begin{figure}[t]
\centering{\includegraphics[width=3.3 in]{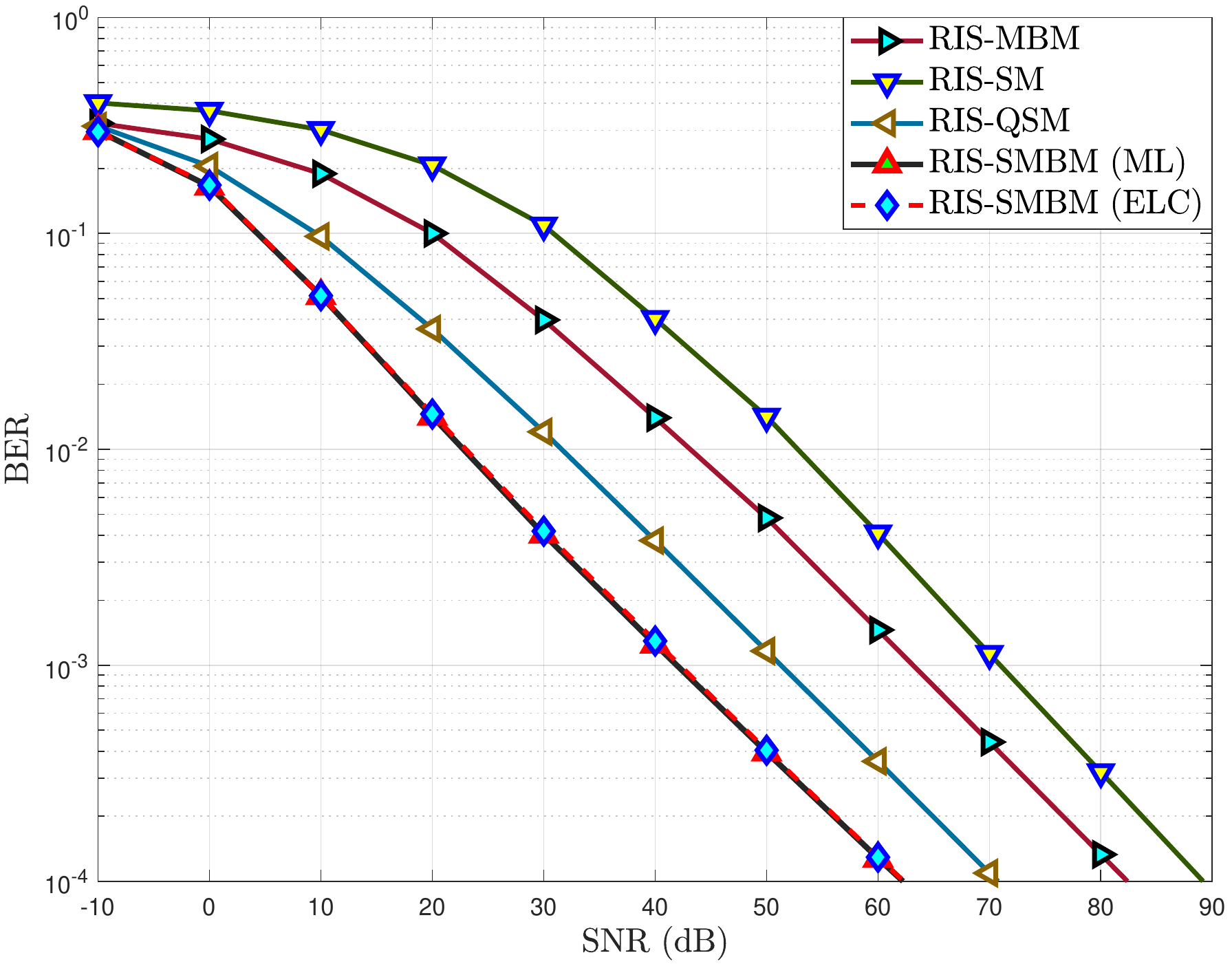}}
	\caption{Simulation performance comparisons of the RIS-SMBM, RIS-MBM, RIS-SM, and RIS-QSM systems while $N=256$ and $\eta=10$ for all systems. }
	\label{SMBMRIS_benchmarks_n10} 
\end{figure}

Fig. \ref{SMBMRIS_benchmarks_n10} presents the performance comparison of the RIS-SMBM system with RIS-QSM, RIS-MBM, and RIS-SM. Also, ($M=128$, $n_T=4$, $m_{rf}=1$), ($M=16$, $m_{rf}=6$), ($M=4$, $n_{T}=256$), and ($M=4$, $n_{T}=16$) are the values of the system parameters for RIS-SMBM, RIS-MBM, RIS-SM, and RIS-QSM systems, respectively.  It is observed that the RIS-SMBM system provides better error performance than RIS-MBM and RIS-SM systems when the spectral efficiency $\eta=10$ bits/s/Hz and $N=256$. RIS-SMBM system has $8.53$ dB, $20.15$ dB, and $26.94$ dB SNR gain according to RIS-QSM, RIS-MBM, and RIS-SM systems, respectively. In addition, similar to the results in Fig. \ref{SMBMRIS_benchmarks_n8}, the BER performances of the RIS-SMBM (ML) and RIS-SMBM (ELC) are pretty similar in Fig. \ref{SMBMRIS_benchmarks_n10}.

\begin{figure}[t]
\vspace{0.233em}
\centering{\includegraphics[width=3.7 in]{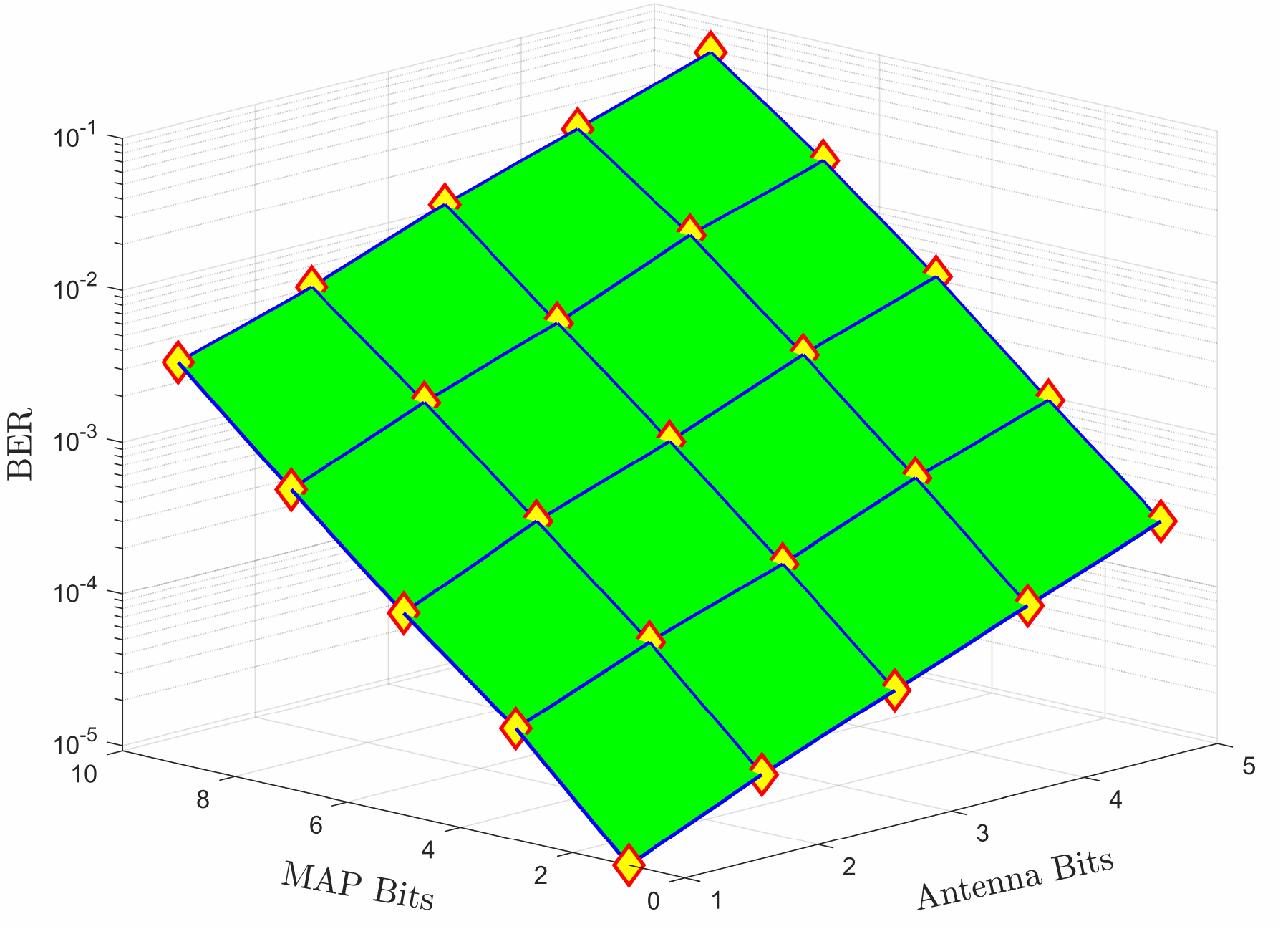}}
	\caption{BER performance of the RIS-SMBM system for $M=4$, $N=32$, and $\text{SNR}=80$ dB. }
	\label{SMBMRIS_3D_NT} 
\end{figure}

Three-dimensional BER performance depending on $m_{rf}=(1, 3, 5, 7 ,9)$ and $n_T=2, 4, 8, 16, 32$ parameters is given in Fig. \ref{SMBMRIS_3D_NT}. Where the number of MAP bits represents the number of bits carried in the MAPs indices formed according to the on/off state of the RF mirrors. In the performance results, instead of the values of the $m_{rf}$ and $n_T$ parameters themselves, the number of bits carried depending on these parameters is considered. It is seen that BER results improve when $m_{rf}$ and $n_T$ values decrease. Therefore, the results show that the minimum BER value occurs for $m_{rf}=1$ and $n_T=2$ and the maximum BER value occurs for $m_{rf}=9$ and $n_T=32$.

\begin{figure}[t]
\vspace{-0.285em}
\centering{\includegraphics[width=3.55 in]{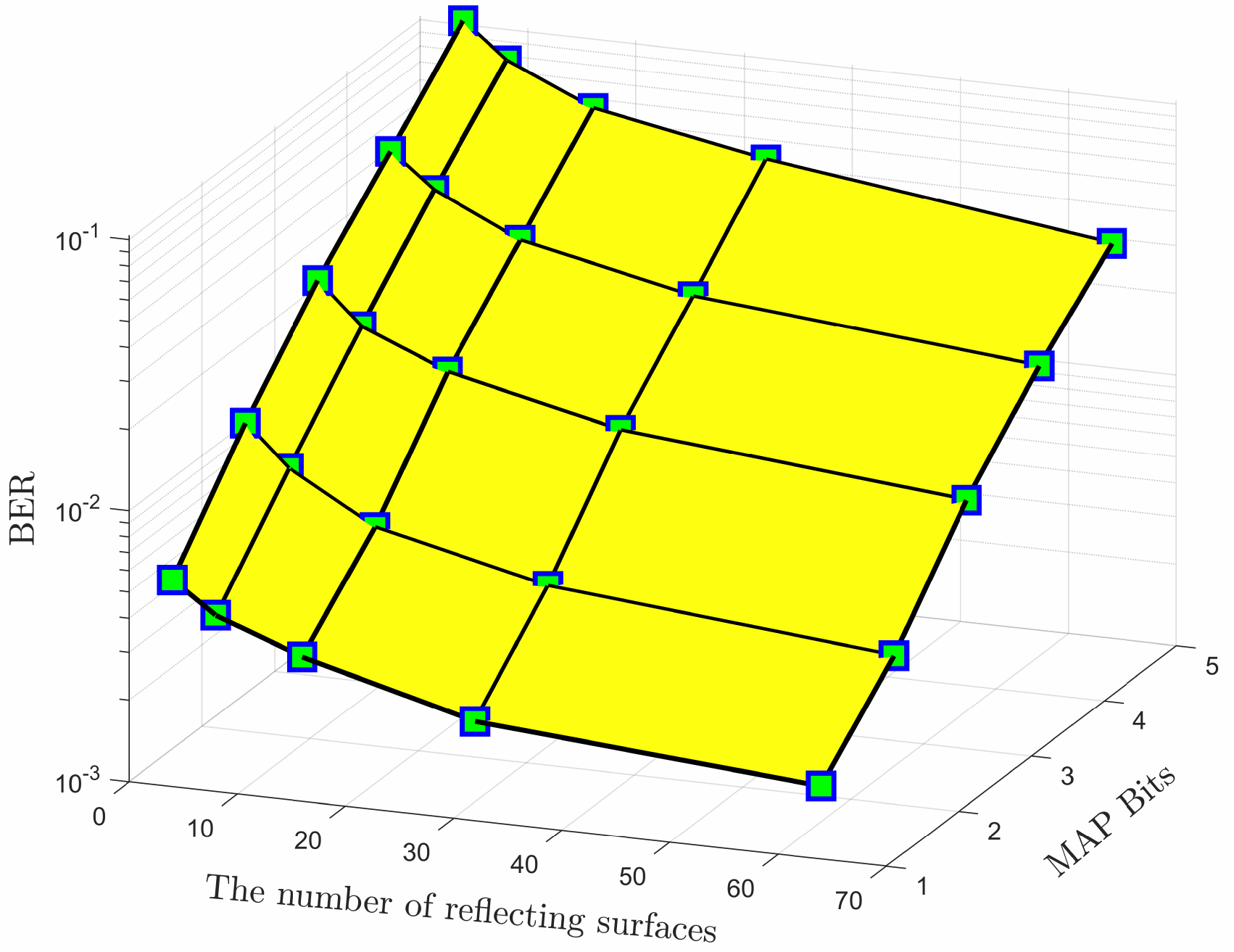}}
	\caption{BER performance of the RIS-SMBM system for $M=4$, $n_T=4$, and $\text{SNR}=40$ dB. }
	\label{SMBMRIS_3D_Ns} 
\end{figure}

Fig. \ref{SMBMRIS_3D_Ns} shows the BER performance of the RIS-SMBM that changes depending on the number of MAP bits and reflecting surfaces. In the simulation in Fig. \ref{SMBMRIS_3D_Ns}, $m_{rf}=(1, 2, 3, 4,5)$ and $N=(4,8,16,32,64)$ values are selected and the effect of these parameters over the error performance is questioned. It has been observed that the error performance gets better as the number of reflecting surfaces increases and the number of MAP bits decreases. As a result, while increasing the number of reflective surfaces improves the system performance, increasing the number of RF mirrors worsens it. However, it should be noted that increasing the number of RF used in the system increases the spectral efficiency of the system. Therefore, the number of RF mirrors and reflective surfaces should be determined in reasonable numbers according to the system's needs.




\section{Conclusions}
In this article, a new system with high performance and high spectral efficiency called RIS-SMBM, which is a combination of RIS, SM, and MBM techniques, is proposed. In the proposed system, data bits are transmitted in the $M$-QAM symbol, active transmit antenna, and active MAP indices. Therefore, the proposed system provides much higher data rates than traditional communication systems. It is shown that the proposed RIS-SMBM system has better BER performance for the same spectral efficiency than RIS-SM, RIS-QSM, and RIS-MBM systems. The error performance change of the RIS-SMBM scheme is observed depending on the RF number $m_{rf}$ and reflecting surface number $N$ parameters. It is observed and discussed that the performance of the proposed system improves as the $N$ number increases and the $m_{rf}$ number decreases. To further reduce the complexity of the RIS-SMBM system, an ELC detector that provides error performance close to the optimal ML detector is proposed. In addition, the computational complexity, energy efficiency, data rate, and throughput of the RIS-SMBM, RIS-SM, RIS-QSM, and RIS-MBM systems are calculated and the results are compared. As a result, the RIS-SMBM system provides a high data rate and high energy efficiency with reasonable complexity compared to counterpart communication systems and transmits data with less error.

\bibliographystyle{ieeetr}

\bibliography{Referanslar}

\end{document}